\documentclass[iop]{emulateapj}
\begin{document}

\title{The Peculiar Pulsar Population of the Central Parsec}
\shorttitle{Galactic center pulsars}
\shortauthors{Dexter \& O'Leary}
\author{Jason Dexter}
\affil{Department of Astronomy, University of California, Berkeley, CA 94720-3411, USA}
\email{jdexter@berkeley.edu}
\author{Ryan M. O'Leary}
\affil{Department of Astronomy, University of California, Berkeley, CA 94720-3411, USA}
\email{oleary@berkeley.edu}

\begin{abstract}
Pulsars orbiting the Galactic center black hole, Sgr A*, would be potential probes of its mass, distance and spin, and may even be used to test general relativity. Despite predictions of large populations of both ordinary and millisecond pulsars in the Galactic center, none have been detected within $25$ pc by deep radio surveys.  One explanation has been that hyperstrong temporal scattering prevents pulsar detections, but the recent discovery of radio pulsations from a highly magnetized neutron star (magnetar) within $0.1$ pc shows that the temporal scattering is much weaker than predicted. We argue that an intrinsic deficit in the ordinary pulsar population is the most likely reason for the lack of detections to date: a ``missing pulsar problem'' in the Galactic center. In contrast, we show that the discovery of a single magnetar implies efficient magnetar formation in the region. If the massive stars in the central parsec form magnetars rather than ordinary pulsars, their short lifetimes could explain the missing pulsars. Efficient magnetar formation could be caused by strongly magnetized progenitors, or could be further evidence of a top-heavy initial mass function. Furthermore, current high-frequency surveys should already be able to detect bright millisecond pulsars, given the measured degree of temporal scattering. 
\end{abstract}

\maketitle

\newcommand{\sgr}{SGR J1745-29 }
\newcommand{\sgrns}{SGR J1745-29}
\newcommand{\nar}{New A Rev.}

\section{Introduction}

The central parsec of the Milky Way contains a large population of
massive O/WR and B stars both as part of a stellar disc
\citep{genzeletal1996,paumardetal2006,luetal2006} and in a cusp
extending towards the Galactic center (GC) black hole, Sgr A* \citep[the
  ``S
  cluster,''][]{schoedeletal2003,ghezetal2005,gillessenetal2009}. The
orbits of these stars around Sgr A*, particularly the star S2/S0-2, measure the black hole mass to be
$\simeq 4\times10^6 M_\odot$ \citep[e.g.,][]{gillessenetal2009}. The detection and timing of pulsars in
similar orbits around Sgr A* would provide precise measurements of the
black hole mass, and potentially detect several other general
relativistic effects
\citep{wexkopeikin1999,krameretal2004,cordesetal2004,pfahlloeb2004}. Based
on the large number of massive stars and a significant
enhancement in the X-ray binary density \citep{munoetal2005}, it is
widely expected that the central parsec hosts a large neutron star
population. \citet{pfahlloeb2004} predicted that $\sim 100-1000$
pulsars should have a semi-major axis $\lesssim 0.02\,$pc from Sgr A*.

It has long been thought that the strong interstellar scattering screen towards the GC, known for decades to blur VLBI images of Sgr A* \citep[e.g.,][]{backer1978,boweretal2006}, prevented the efficient detection of pulsars. If the scattering material is close to the GC, it would strongly broaden pulse profiles, rendering pulsars undetectable at frequencies $\lesssim 1$ GHz \citep{laziocordes1998}. High frequency ($1.5-15$ GHz) pulsar searches have been performed over a range of beam sizes and sensitivities \citep{johnstonetal1995,kleinetal2004,johnstonetal2006,denevaetal2009,macquartetal2010,batesetal2011}. None of these searches have discovered pulsars in the central parsec, and the deepest search at high frequency ($\nu \approx 15\,\rm GHz$) already would be sensitive to a significant fraction of known pulsars \citep{macquartetal2010}.

Recently, \emph{Swift} detected an $X$-ray outburst from the GC \citep{kenneaetal2013}. \emph{NuSTAR} discovered pulsations with a period of $3.76$ s \citep{morietal2013}, and \emph{Chandra} localized the source to a projected separation of only $\simeq 0.1$ pc from Sgr A* \citep{reaetal2013}. Radio pulsations were subsequently detected, allowing for measurements of the dispersion and rotation measures \citep{eatoughetal2013}. The dispersion measure was the highest ever recorded for a pulsar and placed it in the central parsec. Using VLBI observations, \citet{boweretal2013} found the angular broadening for this new highly magnetized neutron star (magnetar), \sgrns, was in excellent agreement with that of Sgr A*, indicating that both sources are behind the same scattering screen, which implies that the screen is uniform on $\gtrsim 0.1$ pc scales. The temporal broadening, however, was much less than previously expected \citep{spitleretal2013}, and similar to pulsars discovered on larger scales ($\sim 50\,\rm pc$). The angular broadening is also comparable to observations of masers in the GC from $\approx 1$ -- $50$ pc \citep{vanlangeveldeetal1992}. These results suggest that past failures to discover pulsars at low frequencies are not simply due to strong interstellar scattering. 

In light of these observations, we assess the central parsec pulsar population. We assume that the scattering screen is far from the GC, and uniformly apply the temporal broadening measurements of \citet{spitleretal2013} to the central parsec. We show the impact of these results on previous observational constraints, and argue that the failure to detect ordinary pulsars in past surveys constitutes a ``missing pulsar problem'' (\S\ref{sec:observ-constr}). In contrast, the detection of even a single magnetar is unexpected, and implies a high magnetar formation efficiency. This efficiency is sufficient to explain the final outcome of all of the massive O/B stars (\S\ref{sec:effic-magn-form}). Millisecond pulsars (MSPs) are not as well constrained, but if present should be detectable in future high-frequency pulsar surveys  (\S\ref{sec:galact-centre-mill}). We discuss possible solutions to the missing pulsar problem and implications of these results in \S\ref{sec:discussion}. A strongly inhomogeneous scattering medium, while plausible, is unsupported by current data. An intrinsic pulsar deficit is a more likely explanation for the lack of detections.

\begin{figure}
\includegraphics[scale=0.65]{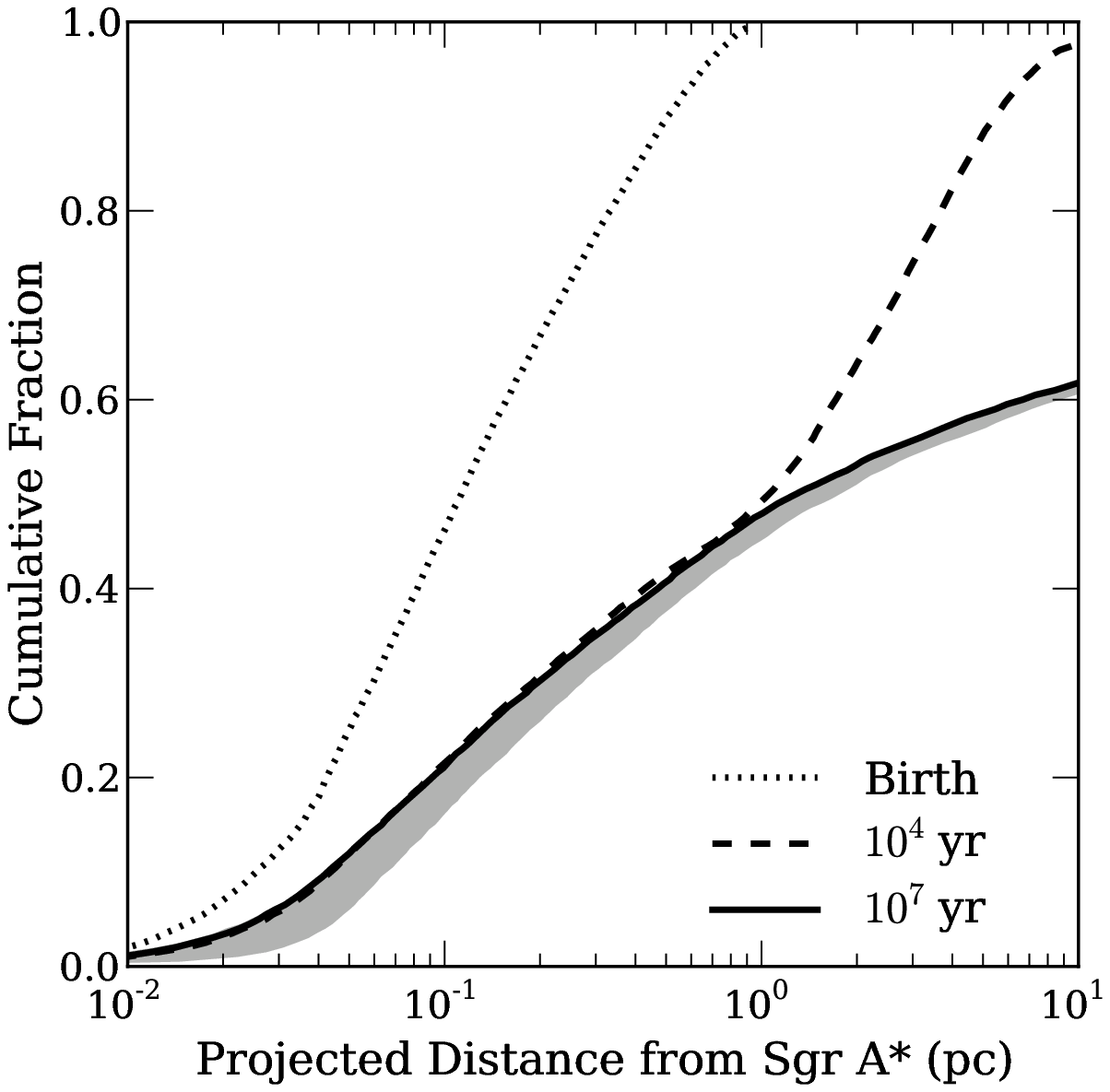}
\caption{\label{retention}The effect of neutron star kicks on isolated
  pulsars born in the central parsec. The cumulative fraction of
  pulsars born in the disk of young stars is shown as a function of projected distance at birth
  \citep[dotted,][]{paumardetal2006}, and at typical ages of
  magnetars (dashed) and ordinary pulsars (solid). In both cases the
  retention fraction of both populations in the central parsec is $\simeq
  0.5$. Nearly all ($\gtrsim 99\%$) of the magnetars remain within the \emph{Swift} 
  field of view ($\approx 60\times60$ pc). The expected cumulative fraction of pulsars within $\lesssim 0.1$ pc depends on the assumed orientation (as shown by the shaded region) and inner edge of the stellar disc.}
\end{figure}
\begin{figure}
\includegraphics[scale=0.65]{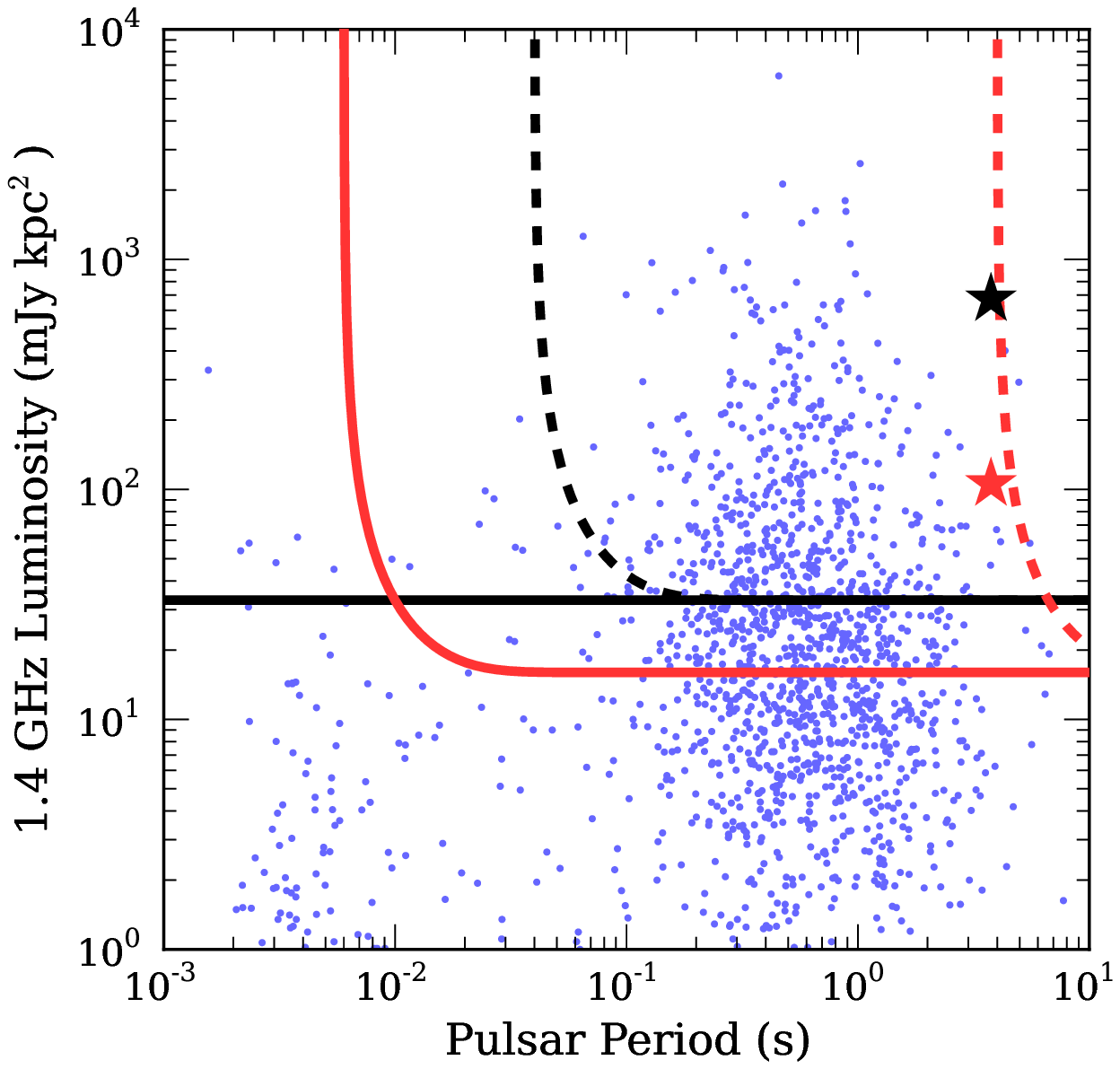}
\caption{\label{constraints}Observational constraints on the GC pulsar population. The dots show the luminosity vs.\ period of known pulsars \citep{manchesteretal2005}. The lines show the sensitivity limits of past surveys and are cut off at $P \lesssim t_{\rm scat}$ as in \citet{macquartetal2010}, where $t_{\rm scat}$ is the previously assumed \citep[dashed,][]{laziocordes1998} and measured \citep[solid,][]{spitleretal2013} temporal scattering towards \sgrns. The black lines are from the $\nu \simeq 14$ GHz survey of \citet{macquartetal2010}, while the tightest constraints, shown as the red lines, come from the Effelsberg \citep{kleinetal2004} and Arecibo \citep{deneva2010,whartonetal2012} surveys at $5$ GHz. Approximately half of known ordinary pulsars could have been detected by past surveys. The luminosity of \sgr re-scaled to $\nu = 5$ and $14$ GHz (red and black stars) is well above the detection threshold of both surveys. The \citet{macquartetal2010} survey at $14$ GHz should be sensitive to pulsars down to the shortest known periods.}
\end{figure}
\section{Young pulsars in the central parsec}
\label{sec:observ-constr}

Pulsars are rotating, magnetized neutron stars formed from the collapse of massive stars ($8-20 M_\sun$) like those in the GC. Pulsars spin down over time, with a characteristic age,

\begin{equation}
T \equiv P / \dot{P} \sim 10 \left(\frac{B}{10^{12} G}\right)^{-2} \left(\frac{P}{1 \hspace{2pt} \rm s}\right)^2 \rm Myr,
\end{equation}

\noindent where $B$ and $P$ are the magnetic field strength and rotation period, scaled to typical values of known pulsars (referred to here as ``young'' or ``ordinary'' pulsars). Magnetars with strong surface fields ($B \gtrsim 10^{14}$ G) spin down much more rapidly and have correspondingly shorter lifetimes ($T \sim 10^4$ yr). In either case, eventually the spin period becomes so long that pulsar emission turns off (the ``death line''). 

The expected total number of young pulsars in the central parsec, $N_{\rm p}$, can be estimated from the number of massive stars and the relative ages of the stars ($\simeq 6$ Myr) and pulsars \citep[][]{pfahlloeb2004}:

\begin{equation}\label{eq:2}
N_{\rm p} \sim 500 \hspace{2pt} \epsilon_p \hspace{2pt} f_{\rm ret} \hspace{2pt} N_{*, \rm GC},
\end{equation}

\noindent where $N_{*, \rm GC} \equiv N_* / 300$, and $N_* \simeq 300$ is a lower limit to the total number of massive stars in the central parsec \citep[e.g.,][spectroscopically identified 177 WR/O/B stars in the central parsec with $< 50\%$ completeness.]{bartkoetal2010}. The parameter $\epsilon_p$ is the pulsar formation efficiency from massive stars, and $f_{\rm ret}$ is the fraction of pulsars which are retained in the inner parsec after their natal kicks.

\subsection{Pulsar kicks}
\label{sec:isol-puls-kicks}

Pulsars with typical kick velocities, $\gtrsim 100 \rm km / s$, can escape the central parsec and deplete the pulsar population. Figure~\ref{retention} shows the expected spatial distribution of neutron stars after $10^7 (10^4)$ yr, assuming that they are kicked from a circular disk of stars similar to that observed in the central parsec \citep{paumardetal2006}  with three-dimensional kick velocities following the double-exponential distribution of \citet{fauchergiguerekaspi2006}. Orbits were integrated using the package {\sc galpy}\footnote{http://code.google.com/p/galpy/}, with a model potential that reproduces the circular velocity of stars from Sgr A* through the Milky Way halo. Approximately half of the isolated pulsars and magnetars escape the inner parsec within their lifetime. Nearly all magnetars remain within the central ten parsecs, within the field of view of the \emph{Swift} $X$-ray telescope. Roughly $20\%$ of the ordinary pulsars can be found outside the central parsec, where they formed, but within $30\,$pc, within the field of view of some previous pulsar surveys. 

However, \citet{whartonetal2012} point out that observations of globular clusters find large pulsar populations despite escape velocities much smaller than the average pulsar kick velocity. This is presumably due to the larger escape velocity needed to escape binary systems, combined with high binary fractions in dense stellar environments. Similar effects could  weaken the impact of neutron star kicks in the GC. We conservatively adopt $f_{\rm ret} = 0.5$ in the following, but this is likely a lower limit.

\subsection{The missing pulsar problem}
\label{sec:mpp}

The number of observable pulsars in the GC depends on the total number of pulsars (Eq. \ref{eq:2}), as well as the survey parameters, pulsar luminosity and period distributions, and the fraction beamed towards us (assumed to be $10\%$ throughout):

\begin{equation}
\label{eq:1}
N_{\rm obs} \simeq 50 f_{\rm ret} f_{\rm det} \epsilon_p N_{*, \rm GC},
\end{equation}

\noindent where $f_{\rm det}$ is the detection efficiency of ordinary pulsars in each survey. Past surveys estimated low values of $f_{\rm det}$ based on the large expected temporal scattering towards the GC. We re-assess the detection efficiencies in light of the measured temporal broadening to \sgr \citep{spitleretal2013}. 

Figure~\ref{constraints} shows the luminosities and periods of known pulsars \citep{manchesteretal2005}, as well as the limits from the deepest surveys at $5$ \citep{kleinetal2004,deneva2010} and $14$ \citep{macquartetal2010} GHz. The previously assumed constraints are shown as dashed lines, while the solid lines use the measured degree of temporal broadening towards the GC. The $5$ GHz results provide the strongest constraint on the ordinary pulsar population, while the $14$ GHz survey may be sensitive to bright MSPs (\S\ref{sec:galact-centre-mill}). Similar results are obtained using the sensitivity of the $8.7$ GHz survey of \citet{deneva2010}.

We estimate $f_{\rm det}$ as the fraction of known ordinary pulsars ($P > 40$ ms) with luminosity larger than the detection threshold of each survey. The detection thresholds are extrapolated to $1.4$ GHz assuming a spectral index of $-1.7$ \citep{krameretal1998}. We find consistent results by instead using the observed spectral indices to extrapolate the pulsar fluxes to the frequencies of the surveys. We roughly account for possible Malmquist bias in the pulsar distribution by doing the estimate separately for all and the closest $50\%$ of pulsars. The result is $f_{\rm det} = 0.5 \pm 0.2$ at $\nu \simeq 5$ GHz and $f_{\rm det} = 0.3 \pm 0.2$ at $\nu \simeq 14$ GHz. 

For $f_{\rm det} \simeq 0.5$, $f_{\rm ret} \ge 0.5$ estimated above, and $\epsilon_p \approx 1$ as expected, the deepest survey should have detected $N_{\rm obs} \approx 10$ ordinary young pulsars. Given an expectation value of $10$, it is statistically improbable ($p \sim 10^{-5}$) to have no detections. Although the systematic uncertainties in this estimate are larger than the statistical ones, they are likely insufficient to explain the discrepancy ($p \sim 10^{-3}$ for the  smallest value of $f_{\rm det}$ at $5$ GHz). This is especially true given our conservative choices for the retention fraction and total number of massive stars. A simple alternative estimate for the expected number of detectable pulsars can be made by scaling the number of observed pulsars ($\sim 2000$) to the fraction of Galactic supernovae that occur in the GC ($\sim 1\%$): $N_{\rm obs} \sim 20 f_{\rm det}$, consistent with the above. 

We conclude that there is a missing pulsar problem in the GC. At $1 \sigma$, an upper limit to the total number of young pulsars is $N_p \lesssim 20$. An alternative view of this constraint is as an upper limit to the pulsar formation efficiency of $\epsilon_p \lesssim 0.1$. That is, $\gtrsim 90\%$ of massive stars in the GC do not form ordinary pulsars. We note that young pulsars make up a small portion of the overall population, and so there could be many more old and/or faint pulsars as discussed by \citet{chennamangalamlorimer2013}. This does not change our estimates of $N_{\rm obs}$ or $\epsilon_p$.

\section{Efficient magnetar formation in the central parsec}
\label{sec:effic-magn-form}

In contrast to the missing pulsars, the detection of a young magnetar ($T \sim 10^4$ yr) so close to Sgr A* implies a large population of highly magnetized neutron stars in the central parsec. The expected number of observable magnetars, $N_{\rm m}$ can be estimated in the same way as above, but for the much shorter magnetar lifetime:

\begin{equation}
N_{\rm m} \sim 0.5 \hspace{2pt} \epsilon_m f_{\rm ret} f_{\rm d} \hspace{2pt} N_{*, \rm GC} \left(\frac{B}{10^{14} G}\right)^{-2} \left(\frac{P}{3.8 \hspace{2pt} \rm s}\right)^2,
\end{equation}

\noindent where the parameters are scaled to the properties of \sgr and $f_{\rm d} < 1$ is the detection fraction of magnetars. We would not have expected to see a magnetar unless magnetar formation is efficient ($\epsilon_m \simeq 1$).

\section{Millisecond pulsar searches}
\label{sec:galact-centre-mill}

Long after a pulsar passes the death line it may accrete mass from a stellar companion, and be rejuvenated in an $X$-ray binary (XRB) as an MSP. The process of forming MSPs is greatly enhanced in dense environments by dynamical encounters between neutron stars and binary stars \citep{verbunthut1987}.  Indeed, old, massive star clusters have some of the richest populations of MSPs and XRBs. \citet{fauchergiguereloeb2011} found that the encounter rate and stellar mass in the GC is similar to the globular cluster Terzan 5, which has a large population of MSPs. They estimated that the GC population could be as large as $\sim 1200$ total MSPs, accounting for its higher pulsar retention fraction. A 
large population of MSPs in the GC is further supported by the overabundance of XRBs found near Sgr A* \citep{munoetal2006}.
To date this population has remained mostly unconstrained. \emph{Fermi} observations are consistent with up to $\approx 10^3$ MSPs within the central $30$ pc \citep{gordonmacias2013}.

Even the relatively small degree of temporal scattering measured by  \citet{spitleretal2013} precludes the efficient detection of MSPs at $\nu \lesssim 8$ GHz.  However, the \citet{macquartetal2010} survey at $\nu = 14\,$GHz could detect the most luminous $\approx 4$ -- $24\%$ of known MSPs. If the GC MSP population were the same as that of Terzan 5, 2 MSPs should have been detected by the \citet{macquartetal2010} survey. 

While the number of MSPs may be enhanced in the GC by dynamical encounters, the estimates typically assume a large population of old neutron stars with low magnetic fields.  If all the neutron stars are magnetars rather than ordinary pulsars, significant field decay would be required to allow spin up to millisecond periods.

\section{Discussion}
\label{sec:discussion}

Despite predictions of a large pulsar population in the central parsec and several deep radio surveys, no ordinary pulsars have been detected to date. Given the recent measurements of weak temporal broadening of the pulses of the magnetar \sgr \citep{spitleretal2013}, previous surveys should have seen $\approx 10$ ordinary pulsars in the central parsec (\S\ref{sec:observ-constr}). We have argued that this discrepancy constitutes a ``missing pulsar problem'' in the Galactic center. The lack of pulsars could be caused by a low formation efficiency, a low detection efficiency, or a low retention fraction of ordinary pulsars. Here we address these possibilities. In particular we argue that a strong but inhomogeneous scattering medium that lowers the detection efficiency of surveys is not the most natural explanation for the missing pulsars.

\emph{A patchy scattering screen:} We have assumed a uniform scattering screen across the inner parsec, as shown to fit both the temporal and angular broadening of \sgr \citep{boweretal2013}. An alternative explanation suggested by \citet{spitleretal2013} is that the discovery of a rare, short-lived magnetar is evidence for a large population of ordinary pulsars as well. These pulsars may not have been seen if the small degree of temporal broadening of \sgr is explained as a lucky line of sight through a much stronger but inhomogeneous scattering screen. 

However, multiple lines of evidence suggest that the angular and temporal scattering may be roughly uniform on scales of tens of parsecs.  \citet{boweretal2013} found excellent agreement in the angular broadening properties of \sgr and Sgr A*, suggesting that the screen has a coherence length $\gtrsim 0.1$ pc. On larger scales, pulsars detected within $\simeq 40$ pc have temporal scattering consistent with or less than that of \sgr \citep{johnstonetal2006,denevaetal2009detect}. Similar angular broadening is seen in many OH/IR stars at separations $\approx 1$ -- $ 50 \rm pc$ \citep{vanlangeveldeetal1992}. Furthermore, the extragalactic source G359.87+0.18 \citep{lazioetal1999} has an angular size consistent with Sgr A* at $1$ GHz, as would be expected for a uniform but distant screen. Since all of the data can be explained by a uniform but distant screen, and in the absence of direct evidence for strong temporal scattering, an intrinsic deficit in the ordinary pulsar population is the more natural interpretation of the lack of detections in surveys to date.

\emph{Detection efficiency biases:} If the luminosity and period distributions of Galactic center pulsars are similar to those detected elsewhere in the Galaxy, the detection efficiency from previous surveys should be high (\S\ref{sec:observ-constr}). However, this distribution is subject to selection effects, and the true distribution could be biased to fainter pulsars. We estimate that this possible bias could contribute at the factor of a few level by separately considering only relatively nearby pulsars, which is probably insufficient to explain the lack of detections. A simple empirical estimate based on the fraction of Galactic supernovae occurring in the GC also gives consistent results, and is not subject to this bias. 

\emph{Neutron star kicks:} We found that $\simeq 50\%$ of the young pulsars could be expelled from the central parsec from natal kicks (\S\ref{sec:isol-puls-kicks}). This is likely an upper limit, as indicated by the large populations of pulsars in globular clusters, which have much smaller escape velocities. Neutron star kicks cannot explain the missing pulsar problem.

\emph{Peculiar star and pulsar formation:} The detection of even a single magnetar so close to Sgr A* is unexpected, and implies order unity efficiency for forming magnetars from massive stars in the Galactic center. This implied high magnetar efficiency is one possible explanation for the missing pulsar problem: massive stars in the central parsec may form short-lived magnetars rather than long-lived ordinary pulsars. Efficient magnetar formation could be explained by an unusual progenitor population. 

There is some evidence that the IMF in the central parsec is top-heavy \citep{nayakshinsunyaev2005,bartkoetal2010,luetal2013}, and several magnetars are associated with massive stellar progenitors \citep[$> 40 \rm M_{\sun}$,][compared to $8-20 M_{\sun}$ typically assumed for neutron stars]{figeretal2005,gaensleretal2005,munoetal2006}. Stars in the central parsec may also be highly magnetized (e.g., S0-2/S2 \citealt{martinsetal2008}, see also the hypervelocity stars, \citealt{przybillaetal2008,brownetal2013}). Magnetars could also be formed from amplification of the field of a highly magnetized progenitor during core collapse \citep{ferrariowickramasinghe2006}. Both a top-heavy IMF and strong magnetic fields could be caused by rapid star formation in the dense regions near Sgr A* \citep{bonnellrice2008,hobbsnayakshin2009}.

The star formation event leading to the currently observed massive stars in the central parsec was too recent to have formed ordinary pulsars. Another explanation for missing young pulsars is that past star formation events produced fewer (by $\approx 90\%$) massive stars (e.g., from a different IMF). A very recent major merger between Sgr A* and another massive black hole could also have removed a significant fraction of the old stars and pulsars from the central parsec. Finally, it is possible that the ordinary pulsars are present, but that the pulsar emission is suppressed, e.g., by interactions of the neutron star magnetosphere with the GC environment over the pulsar lifetime.

Whatever the explanation, missing pulsars and efficient magnetar formation constitute evidence that neutron star formation and evolution are substantially modified in the high density regions near a massive black hole.

\begin{acknowledgements}
The authors are listed in alphabetical order. J.D.\ and R.M.O.\ contributed equally to the writing
and calculations presented in this study.
We thank G.\ Bower, M.\ Kerr, C.\ Law, and A.-M.\ Madigan for helpful discussions related to this work, and the participants of IAU Symposium 303 for coining the phrase ``missing pulsar problem.''  We thank the referee, S. Nayakshin, for helpful suggestions, and J.\ Bovy for publicly releasing the software package {\sc galpy}, which we used in this work.  
\end{acknowledgements}

\bibliographystyle{apj}


\end{document}